\documentclass[pra,amsmath,amsfonts,amssymb,twocolumn,showpacs]{revtex4}
\usepackage{hyperref}
\usepackage{graphicx}

\begin{document}
\newcommand{\eq}[1]{Eq.~(\ref{#1})}

\title{Nonstationary electromagnetics of controllably dispersive media}
\author{Douglas H. Bradshaw}
\email{bradshaw@lanl.gov}
\author{Michael D. Di Rosa}
\affiliation{Los Alamos National Laboratory, Los Alamos, NM 87545 USA}

\begin{abstract}
Recent experiments have demonstrated that it is possible to alter the dispersion of a medium without significantly altering its absorption or refractive index and that this may be done while a wave propagates through the medium.  This possibility opens up a new set of potential experiments to the field of nonstationary optics.  We consider the basic kinetics of waves propagating through a medium whose group and phase velocities are a function of position and time.  We compare the dynamics of waves propagating through two homogeneous media, one a nondispersive medium with a time dependent phase velocity, and one a dispersive medium with a time dependent group velocity and show that new dynamic effects accompany new kinetic ones.
\end{abstract}
\maketitle

\begin{section}{Introduction}\label{se:intro}
The terms \emph{nonstationary}, \emph{inhomogeneous}, and \emph{dispersive} describe media whose properties vary with time, position, and frequency, respectively.  In this paper, we consider transformations wrought upon waves as they propagate through nonstationary, inhomogeneous media with variable dispersion.  Our interest in this topic is motivated by experimental advances in the manipulation of dispersion through nonlinear optics.  Recent experiments have demonstrated that it is possible to achieve either strong normal dispersion (in the case ``slow-light media \cite{Hau-Harris-Dutton-Behroozi-1999,Liu-Dutton-Behroozi-Hau-2001}'') or strong anomalous dispersion (in the case of ``superluminal media \cite{Wang-Kuzmich-Dogariu-2000}'') in nearly transparent media using nonlinear optical effects.  In either case, the strength of the dispersion, as experienced by a probe beam, depends on the intensity of one or more controllable auxiliary beams.  If auxiliary intensities are changed while the probe beam is in transit, then the dispersion, as perceived by the probe beam, is time dependent.  These media, which we will refer to collectively as controllably dispersive media, thus form a newly accessible class of nonstationary media.  Although the nonstationarity of a slow light medium has been used experimentally to adiabatically transform a ``slow light'' pulse to a ``stopped'' one \cite{Liu-Dutton-Behroozi-Hau-2001,Ginsberg-Garner-Hau-2007}, an explicit connection between this application and the field of nonstationary electromagnetics has not, to our knowledge, been made.

The field of nonstationary electromagnetics, although much less developed than its inhomogeneous counterpart, is now a venerable one.  In 1958, Morgenthaler introduced propagation equations for electromagnetic waves in an isotropic nondispersive medium whose permittivity and permeability were allowed to vary with time (but not with space) \cite{Morgenthaler-1958}.  Although the propagation equations are not generally analytically soluble, he found useful solutions assuming, on the one hand, a step functional time dependence and, on the other hand, an adiabatic time dependence.  He found, among other things, that the frequency of a wave would vary with the permittivity and permeability in such a way that wavelength would be preserved.  

Over the next two decades, much of the progress of the field was in the Soviet Union, with a particular focus on the behavior of light in plasmas.  We refer the reader to Stepanov and colleagues \cite{Stepanov-1993,Kratsov-Ostrovsky-Stepanov-1974} for an introduction to this literature and highlight here two aspects of it.  First, the two simplified cases (step function and adiabatic changes) used by Morgenthaler continued to be used extensively, and were generalized from a purely nonstationary picture to one that allowed inhomogeneities to propagate at a fixed velocity according to the traveling wave law ($f=f(x-vt)$) \cite{Stepanov-1993, Stepanov-1969}.  This picture allows a single framework to unite the purely nonstationary case ($v\rightarrow \infty$) \footnote{Here we note a basic characteristic of nonstationary media: there is no maximum speed because changes in a medium at two different positions need not be causally connected to each other.}, a 1-dimensional purely inhomogeneous case ($v=0$), and the case of an inhomogeneity (for example, an ionization front) moving at any velocity in between these two extremes.  
Second, a distinction between ``kinetic'' and ``dynamic'' phenomena was found to be useful in differentiating between those results of the theory that are general to all linear wave phenomena and those that are specific to particular media \cite{Stepanov-1969, Stepanov-1993, Sorokin-Stepanov-1971}.  We employ both of these concepts in this paper.

The advent of the laser, and particularly the pulsed laser, introduced a new way to dynamically alter the characteristics of a medium.  Although they were apparently unaware of previous work in nonstationary electromagnetics, a few authors noted the possibility of using light to alter the phase velocity of propagating electromagnetic waves.  In 1977, Lampe, Ott, and Walker noted that an ionizing laser might be swept across a gas to create a superluminal ionization front that could interact with a microwave pulse in a nonstationary fashion \cite{Lampe-Ott-1977}.  In 1988, Wilks, Dawson, and Mori examined the problem of a wave propagating through a medium which is then quickly ionized by a high-intensity ultrashort pulse \cite{Wilks-Dawson-Mori-1988}.  Large frequency shifts were soon realized for waves reflecting off of a relativistically propagating ionization front \cite{Savage-Joshi-Mori-1992}.  These works founded a sub-literature surrounding nonstationary effects in plasmas \cite{Berezhiani-Mahajan-Miklaszewski-1999, Hashimony-Zigler-Papadopoulos-2001,Geltner-Avitzour-Suckewer-2002,Avitzour-Geltner-Suckewer-2005, Shvartsburg-2005,Avitzour-Shvets-2008}.  We point the reader to Shvartsburg, who reviewed some of the recent work on nonstationary effects in plasmas \cite{Shvartsburg-2005}.  We also highlight the work of Avitzour and Shvets, whose 2008 paper relies on controlled dispersion in nonstationary media.  They proposed a method for altering the spectral width of a pulse without changing its central frequency using controlled dispersion in a nonstationary magnetized plasma \cite{Avitzour-Shvets-2008}.  We will look at this method from a kinetic perspective in the next section.

In parallel with the more recent work on nonstationary plasmas, several authors have explored electromagnetic propagation through nonstationary media abstractly, in the tradition of Morgenthaler \cite{Dodonov-Klimov-Nikonov-1993, Biancalana-Amann-Uskov-Oreilly-2007,Budko-2009}.  Dodonov, Klimov, and Nikonov studied the quantization of a linear, nonstationary medium and were able to quantitatively relate temporal changes in dielectric permittivity to photon generation \cite{Dodonov-Klimov-Nikonov-1993}. Biancalana, Amann, Uskov, and Oreilly treated a 1-dimensional nondispersive nonstationary medium by introducing a transmission matrix for moving interfaces.  Using this matrix they generalized Bragg reflection to propagating interfaces and showed that temporal periodicity leads to k-vector bandgaps just as spatial periodicity leads to frequency bandgaps \cite{Biancalana-Amann-Uskov-Oreilly-2007}.  Budko has found an analogy between a nonstationary nondispersive medium and an expanding Universe \cite{Budko-2009}.

One surprising fact about recent works, both those focusing on nonstationary plasmas \cite{Lampe-Ott-1977,Wilks-Dawson-Mori-1988,Savage-Joshi-Mori-1992,Berezhiani-Mahajan-Miklaszewski-1999, Hashimony-Zigler-Papadopoulos-2001,Geltner-Avitzour-Suckewer-2002,Avitzour-Geltner-Suckewer-2005, Shvartsburg-2005,Avitzour-Shvets-2008} and those dealing with abstract, nondispersive, linear media \cite{Dodonov-Klimov-Nikonov-1993, Biancalana-Amann-Uskov-Oreilly-2007,Budko-2009}, is that they make no reference to the substantial Soviet literature from the 1960s and 1970s.  One consequence of the disjointedness of which this fact is symptomatic is that basic principles have had to be rediscovered, multiple times, in incidental and typically incomplete ways.  For example, a basic principle of wave propagation in nonstationary linear media is the importance and generality of kinetic effects. Simple spatial and temporal symmetries play an important role in determining the effects of nonstationary propagating waves.  Aspects of these symmetries have been rediscovered and used in many contexts \cite{Morgenthaler-1958, Stepanov-1969, Wilks-Dawson-Mori-1988,Biancalana-Amann-Uskov-Oreilly-2007,Avitzour-Shvets-2008,Budko-2009}.

We hope that this paper, in addition to providing a simple theoretical basis for treating a new class of nonstationary dispersive media, will demonstrate the utility of explicitly recognizing kinetic constraints and of maintaining a clear distinction between kinetic and dynamic aspects of wave behavior.  Section~\ref{Sec:symmetry} is dedicated to an exploration of the interplay between dispersion, nonstationarity, and inhomogeneity from a strictly kinetic perspective.  We introduce three basic kinetic relationships representing three basic preserved symmetries. Along with appropriate dispersion relationships, they allow for the derivation of all the subsequent kinetic results.  Although these relationships have not, to our knowledge, been explicitly presented together before, they have been implicit in many previously derived results.  We show that they lead to Snell's Law for a motionless interface and to appropriate Doppler shifts for a moving one, that they lead to the preservation of wavelength for homogeneous nonstationary media and to the preservation of frequency in inhomogeneous stationary media, how they give rise to Biancalana \emph{et al.}'s ``generalized frequency'' and give insight to Avitzour and Shvets' proposal for compressing pulse spectrum without altering carrier frequency.  We also find some effects that are described here for what may be the first time.  For example, we find that dispersion modulates the frequency response of a wave to temporal changes in the refractive index and that Doppler reflections may lead to large changes in pulse bandwidth when the moving interface approaches the group velocity.  We emphasize that although we have derived these results with controllably dispersive nonlinear optical media in mind, the results apply to any type of wave propagating through any linear medium.

Unlike kinetic effects, the dynamic aspects of wave behavior depend on the specific microscopic details of interactions between nonstationary media and fields.  In Section~\ref{Sec:comp}, we develop boundary conditions appropriate to an idealized version of the nonlinear optical media currently used to achieve exotic dispersion.  Unsurprisingly, we find that these boundary conditions are fundamentally different from those posited for nondispersive media.  We compare pulse transformations wrought by a changing phase velocity in a specific, homogeneous, nondispersive medium (originally described by Morgenthaler \cite{Morgenthaler-1958}), with those wrought by a change in group velocity in our idealized dispersive medium.  We note many interesting similarities and differences in their effects on twenty key quantities such as energy, momentum, and photon density.

\end{section}

\begin{section}{Kinetics of dispersive, inhomogeneous, nonstationary media}\label{Sec:symmetry}

\begin{figure}
	\centerline{\includegraphics[width=8.5cm]{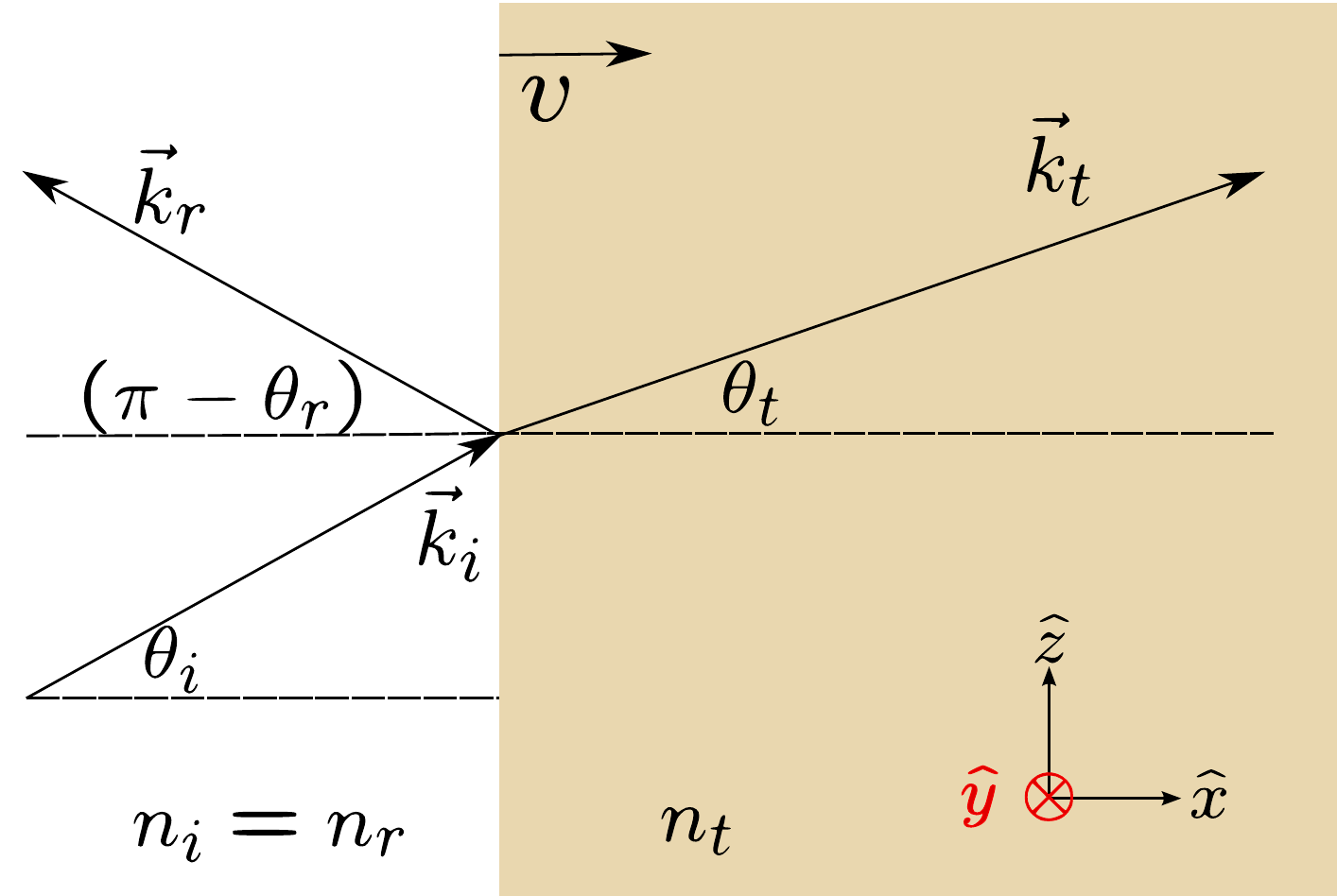}}
	\caption{\label{Fig:interface} An incident plane wave with k-vector $\vec k_i$ interacts with a planar interface.  This results in a refracted wave with k-vector $\vec k_r$ and a transmitted wave with k-vector $\vec k_t$.  For a non-dispersive and isotropic medium, the refractive index for the reflected wave ($n_r$) is the same as that for the incident wave ($n_i$).}
\end{figure}

Consider a plane wave incident on an infinite planar interface, as drawn in Fig.~\ref{Fig:interface}.  Its wave vector and frequency describe periodic translational symmetries in 3 dimensions of space and 1 of time.  A planar interface breaks symmetry in only one dimension, i.e., the one parallel to its normal.
If an interface is motionless, then its normal is purely spatial; in the geometry of Fig.~\ref{Fig:interface}, we could write $\hat n= \hat x$.  The normal of a moving interface in 3+1 dimensions is partially rotated into time; if we introduce the temporal unit vector $\hat{ct}$ and a temporal rotation angle $\phi$, then we may write $\hat n = \cos\phi \,\hat x+\sin\phi\,\hat{ct}$.  The angle $\phi$ is related to the velocity $v$ and the normalized velocity $\beta$ by
\begin{equation}\label{Eq:phi}
 \frac{v}{c}=\beta=\tan\phi.
\end{equation}
All the other translational symmetries of the incident plane wave are preserved under reflection and transmission.
Thus,
\begin{equation}\label{Eq:ky}
k_{yi}=k_{yr}=k_{yt},
\end{equation} 
\begin{equation}\label{Eq:kz}
k_{zi}=k_{zr}=k_{zt},
\end{equation} 
and
\begin{equation}\label{Eq:ktx}
\frac{\omega_i}{c}-k_{xi}\beta=\frac{\omega_r}{c}-k_{xr}\beta=\frac{\omega_t}{c}-k_{xt}\beta,
\end{equation}
where the geometry is again defined by Fig.~\ref{Fig:interface}.
These three equations, combined with appropriate dispersion relationships, provide a basis for the kinetic aspects of wave behavior.

Equations (\ref{Eq:ky}), (\ref{Eq:kz}), and (\ref{Eq:ktx}) represent quantities that are unchanged by a planar interface.  Extending this concept, we see that if these quantities are not altered by a single interface, they cannot be altered by multiple similar interfaces (``similar'' means here that they share the same velocity and the same normal).  By progressively increasing the number and variety of similar interfaces, we can similarly see that the same quantities are preserved under interaction with any pattern of medium parameters that obeys the traveling wave law, 
\begin{equation}\label{Eq:travel}
f=f(x-vt).
\end{equation}

When the planar disturbance described by the function $f$ in \eq{Eq:travel} is more complicated than a single interface, the description of the wave response in terms of a single incident, reflected, and transmitted wave becomes inadequate.  In order to express conserved quantities appropriately for this more general case, we now introduce the convention that a line appearing over a given quantity means that the quantity is invariant under planar changes in the material parameters.  Using this convention, we extend Equations (\ref{Eq:ky}), (\ref{Eq:kz}), and (\ref{Eq:ktx}) to
\begin{equation}\label{Eq:kyg}
 \overline{k_y},
\end{equation} 
\begin{equation}\label{Eq:kzg}
 \overline{k_z},
\end{equation} 
and
\begin{equation}\label{Eq:ktxg}
 \overline{\frac{\omega}{c}\cos(\phi)-k_x \sin(\phi)}.
\end{equation}
Despite their lack of an explicit equality, we refer these statements equations in acknowledgment of the fact that they are, in fact, compressed equalities.

\subsection{Snell's Law and Doppler shifts}
For a motionless interface, $\beta=0$ and \eq{Eq:ktx} reduces to the statement that frequency is preserved:
\begin{equation}\label{Eq:omega}
 \omega_i=\omega_r=\omega_t.
\end{equation}  
Combining this with \eq{Eq:kz} and the expansion $k_z=n\omega\sin\theta/c$ yields
\begin{equation}\label{Eq:Snell}
 \sin\theta_i=\sin\theta_r=\sin\theta_t,
\end{equation} 
which is Snell's Law (note that Snell's Law applies to the reflected wave as well as the transmitted wave).

If we allow the interface to move at a velocity $v$, frequency is no longer conserved, and Snell's Law will no longer hold \footnote{For a moving object in a vacuum, this is easily solved via Lorentz transformation.  For general nonstationary objects, such a transformation may be costly: isotropic media may be come anisotropic.  In addition, Lorentz transformations cannot be used to simplify the treatment of interfaces moving faster than the speed of light.}.  If both sides of the interface are dispersionless, then a simple extension of Snell's Law obtained by combining \eq{Eq:ktx} and \eq{Eq:kz} may be useful:
\begin{equation}\label{Eq:Snell2}
 \frac{n_t\sin\theta_t}{1-n_t \cos\theta_t \beta}=\frac{n_i\sin\theta_i}{1-n_i \cos\theta_i \beta}.
\end{equation} 
One interesting fact about this formula is that it shows that there are some cases where there two sets of angles and frequencies for the transmitted wave \footnote{Multiple transmitted solutions can correspond to the absence of a reflected solution.  This has been noted in the past for moving interfaces.  See \cite{Lampe-Ott-1977}}.
When dispersion is important, \eq{Eq:Snell2} loses its utility because $\theta_t$ and $n_t$ both become frequency dependent.  We may still find angle, refractive index, and frequency using both relationships
\begin{equation}
 n_t(\omega_t)\omega_t\sin\theta_t=n_i(\omega_i)\omega_i\sin\theta_i,
\end{equation} 
and
\begin{equation}
 \omega_t(1-n_t(\omega_t)\cos\theta_t\beta)=\omega_i(1-n_i(\omega_i)\cos\theta_i\beta),
\end{equation} 
combined with an explicit function for $n_t(\omega_t)$ to calculate the properties of the transmitted wave.  We will explore the effects of dispersion in a simplified case, that of the transmitted wave at normal incidence, in the next subsection.

The relationships for the reflected wave,
\begin{equation}
 n_r(\omega_t)\omega_r\sin\theta_r=n_i(\omega_i)\omega_i\sin\theta_i,
\end{equation} 
and
\begin{equation}\label{Eq:Doppler2}
 \omega_r(1-n_r(\omega_r)\cos\theta_r\beta)=\omega_i(1-n_i(\omega_i)\cos\theta_i\beta),
\end{equation}
combined with an explicitly function for $n_r(\omega_r)$ appear identical to those for the transmitted wave.  However, the behavior is different in this case because the medium that defines $n_r$ is the same as the medium that defines $n_i$ and because the sign of $\cos\theta$ is different in the reflective case.  Numerical investigation suggests that the Doppler effect grows large when $|v_g| \approx v$ for an approaching wall.  In Subsection \ref{ss:Doppler1D} we will show analytically that this is the case for a wave that is normal to the interface.  For now, we note that for normal incidence on a moving interface in a vacuum, $\cos\theta_r=-1$, $\cos\theta_i=1$, and $n_r=n_i=1$.  Then \eq{Eq:Doppler2} gives the standard reflective Doppler shift,
\begin{equation}
 \omega_r=\omega_i\frac{1-\beta}{1+\beta}.
\end{equation}  
Note that in the geometry of Fig.~\ref{Fig:interface} a positive value for $\beta$ means that the interface is receding.

\subsection{Transmission at normal incidence}
In controllably dispersive media, large changes in group velocity may be accompanied by minute changes in phase velocity.  In addition, the phase velocity changes are gradual in the sense that they occur over several carrier frequency cycles.  Under these conditions, reflection is negligible.
At normal incidence, $\cos\theta_i=\cos\theta_t=1$.  Then \eq{Eq:ktxg} simplifies to
\begin{equation}\label{Eq:gen_omega}
 \overline{\frac{\omega}{c}\cos\phi-k\sin\phi},
\end{equation}
where we have removed the subscript $x$ from $k_x$ in acknowledgment of the fact that at normal incidence $k=k_x$.
This mixture of wave vector and frequency is the conserved quantity that Biancalana \emph{et al.} found and referred to as the generalized frequency \cite{Biancalana-Amann-Uskov-Oreilly-2007}.  When $\phi=0$, the medium is stationary and frequency is preserved.  When $\phi= \pm\pi/2$, the medium is homogeneous and wave vector is preserved.  In between these extremes, a mixture between wave vector and frequency is preserved.

The preservation of the generalized frequency is accompanied by a second constraint: 
\begin{equation}\label{Eq:v_p}
 \frac{\omega}{k}=v_p,
\end{equation} 
where $v_p$ is the phase velocity.  When the phase velocity changes, the ratio $\omega/k$ must change as well.  Together, Eqs. (\ref{Eq:gen_omega}) and (\ref{Eq:v_p}) determine the frequency and wave vector responses.  We now look at the impact of dispersion upon these responses.  Consider a medium for which the spectral dependence and the time dependence of the refractive index are independent so that we may apply $n=n_o(x,t)+(\delta n/(\delta \omega))(\omega-\omega_o)$ for frequencies near $\omega_o$.
From
\begin{equation}
 \frac{d}{dn_o}\left(\frac{\omega}{c}\cos\phi-k\sin\phi\right)=0,
\end{equation} 
we obtain
\begin{equation}
 \frac{d\omega}{dn_o}=\frac{\omega \sin\phi}{\cos\phi-n_g \sin\phi},
\end{equation} 
where $n_g=n+\omega \delta n/\delta \omega$.  Since $\tan\phi=v/c$, we can also write
\begin{equation}
 \frac{d\omega}{dn_o}=\frac{\omega v/c}{1-v/v_g},
\end{equation}
which emphasizes the large frequency shifts expected when the interface velocity $v$ approaches the group velocity $v_g$.

\subsection{Preservation of a generalized bandwidth in 1+1 dimensions}\label{sse:generalized-bandwidth}
The group velocity is defined using $v_g=d\omega/dk$.  When group velocity dispersion may be neglected, either because it is small, or because we are limiting ourselves to a narrow spectrum, the approximation
\begin{equation}\label{Eq:domega-dk}
 \frac{\Delta \omega}{\Delta k} \approx v_g
\end{equation}
becomes useful. 
As formalized in Eq. ~(\ref{Eq:domega-dk}), when a narrow band pulse experiences a change in group velocity, either its spectral bandwidth $\Delta \omega$, or its spatial bandwidth $\Delta k$, or both, must change.  Extending Eq. ~(\ref{Eq:gen_omega}), we find a generalized bandwidth,
\begin{equation}\label{Eq:gen-bandwidth}
\overline{\frac{\Delta \omega}{c} \cos\phi-\Delta k \sin\phi},
\end{equation} 
that is preserved when a pulse interacts with an interface moving at the velocity defined by $\phi$ and Eq ~(\ref{Eq:phi}).  \eq{Eq:gen-bandwidth} shows that the division of change between spectral and spatial bandwidth depends on their comparative magnitudes and on the interface velocity.

The preserved generalized bandwidth provides us with a simple tool that we can use to analyze one particular aspect of the stopped light experiment of Liu, Dutton, Behroozi, and Hau \cite{Liu-Dutton-Behroozi-Hau-2001}.  In the Liu experiment, a coupling beam was used to control the group index perceived by a probe beam in real time.  However, this coupling beam had a finite speed and was copropagating with the probe beam.  Because of this, changes to the group index perceived by the probe beam were not instantaneous, but moved at nearly the speed of light.  In their report \cite{Liu-Dutton-Behroozi-Hau-2001}, and in earlier theoretical work \cite{Fleischhauer-Lukin-2000}, the speed of the coupling beam was treated as effectively infinite, because it was so much faster than the slowed advancement of the probe pulse.  Using the preserved generalized bandwidth, we now show that treating the velocity of the coupling beam as essentially infinite was justified.

When the pulse described by Liu \emph{et al.} entered the slow light medium, it crossed a spatial interface, and $\phi$ was $0$.  The preserved quantity associated with this case was $\overline{\Delta \omega/c}$, meaning that spectral bandwidth was preserved.  The spatial bandwidth ($\Delta k$) and the group index ($n_g$) were increased by 7 orders of magnitude, and Eq. ~(\ref{Eq:domega-dk}) remained satisfied.  

While the new foreshortened pulse was propagating through the slow-light medium, the coupling beam was then turned off.  The speed of the coupling beam was approximately $c$, so that $\phi$ was approximately $\pi/4$.  Thus, the preserved generalized frequency was approximately $\overline{(\Delta\omega/c-\Delta k)/\sqrt{2}}$, which appears to distribute changes evenly between spectral and spatial bandwidths.  However, the slow speed of the probe envelope was reflected by the fact that $\Delta k$ was larger than $\Delta \omega/c$ by a factor of 10$^7$.  According to the preservation of the generalized bandwidth, $\Delta \omega/c$ would have gone to zero while $\Delta k$ changed by 1 part in 10$^7$.  This is the size of the error introduced by assuming that the coupling speed was effectively infinite.

The end result of the Liu experiment was a reconstitution of the original pulse.  In the next section we use the concept of preserved generalized bandwidth to show how the pulse may be spectrally compressed or expanded using transmission in controllably dispersive media.

\subsection{Pulse compression/decompression using controlled dispersion}
\begin{figure}
  \centerline{\includegraphics[width=8.5cm]{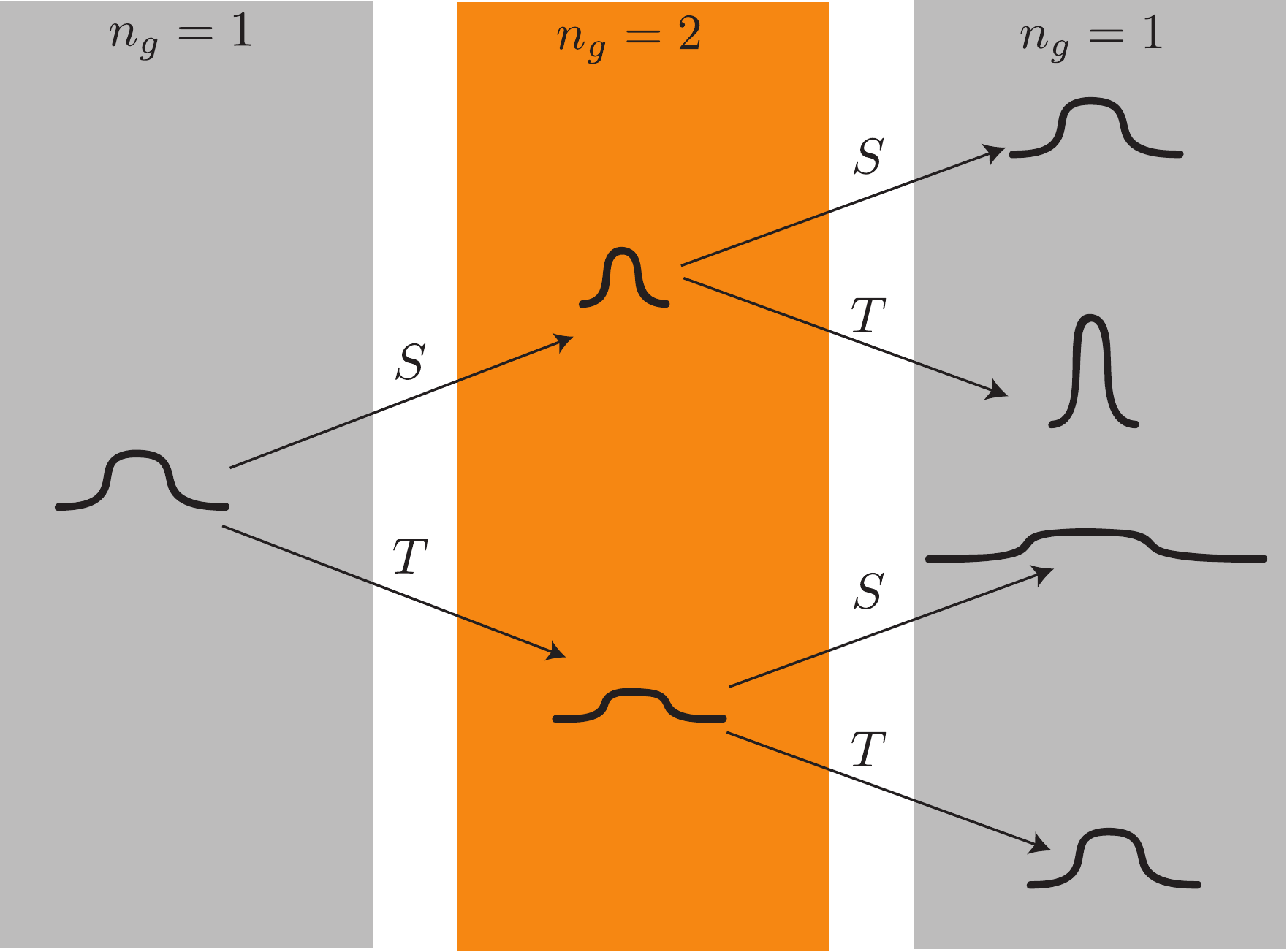}}
  \caption{\label{Fig:pulse-scaling} Field energy density verses spatial extent of a pulse.  S: when a pulse undergoes a spatial change in group index ($n_g$) it scales longitudinally but not temporally.  T: when a pulse undergoes a temporal change in ($n_g$), it changes temporally but not longitudinally.  By mixing spatial and temporal interfaces, it is possible to scale the pulse bandwidth and longitudinal extent without otherwise changing the envelope shape.}
\end{figure}

The generalized bandwidth is conserved across any number of interfaces so long as they have the same velocity.  By mixing interfaces of different velocities, we can change that bandwidth.  In Fig.~\ref{Fig:pulse-scaling}, a pulse begins in a medium with a group index $n_g$ of 1, transitions to one with $n_g=2$, and then back to $n_g=1$.  When both transitions are spatial (S corresponds to a motionless planar interface) or when both are temporal (T corresponds to time-dependent but spatially independent change to the medium--the infinite velocity case) the pulse ends up in its original form.  By mixing spatial and temporal transformations, the pulse can either be compressed or extended.  In this way the pulse bandwidth may be compressed or decompressed while the pulse envelope is scaled longitudinally but otherwise unchanged.

This mechanism for pulse compression and decompression was recently proposed and demonstrated in simulation for a plasma geometries, where the group velocity may be controlled dynamically via an applied magnetic field \cite{Avitzour-Shvets-2008}.  We note here that the Liu experiment demonstrates all the capabilities necessary to perform pulse compression or expansion using nonlinear optics \cite{Liu-Dutton-Behroozi-Hau-2001}.

\subsection{Dispersive modulation of the 1-D Doppler shift}\label{ss:Doppler1D}
In a dispersive medium, Doppler shifts are difficult to predict intuitively because of the interdependence between refractive index and frequency.  Here we find an intuitive formula for the change in frequency in the case where the group velocity differs from the phase velocity, but where the range of frequencies concerning us is sufficiently narrow that we may neglect group velocity dispersion.  Expanding \eq{Eq:ktx} for the case of reflection at normal incidence, we get 
\begin{equation}\label{Eq:1D-Dopper-general}
 \omega_i\left(1-n_i \frac{v}{c}\right)=\omega_r\left(1+n_r \frac{v}{c}\right).
\end{equation}
To derive a Doppler shift, we must introduce an explicit dispersion relationship $n=n(\omega)$.  In general, the resulting equation is analytically intractable though a numerical solution is readily found.  An interesting argument for the physical relevance of the concept of group velocity to Doppler shifts in dispersive media is that if we require that the group velocity be constant with frequency a simple analytical solution results.  Requiring that $n_g$, the group index, remain constant with frequency gives a refractive index of the form $n=n_g+b/\omega$, where $b$ can be any real constant.  Applying this dispersion relationship to Eq.~(\ref{Eq:1D-Dopper-general}) yields the exact expression,
\begin{equation}\label{Eq:Doppler-no-dispersion}
 \frac{\Delta}{\omega}=-2 n \frac{v}{c},
\end{equation}
where $\beta=v/c$, $\Delta=\omega_r-\omega_i$, $\omega_r+\omega_i=2\omega$, and $n=n(\omega)$.
Eq.~(\ref{Eq:Doppler-no-dispersion}) is attractive because of its symmetry but incomplete because it depends on the quantity $\omega$, which it does not resolve.  Rewriting Eq.~(\ref{Eq:Doppler-no-dispersion}) in terms of $\omega_i$ and $\Delta$
\begin{equation}\label{Eq:Doppler-dispersion}
 \frac{\Delta}{\omega_i}=\frac{-2 n(\omega_i) v/c}{1+v/v_g},
\end{equation} 
which clearly reveals the role of dispersion in determining the expected Doppler shift.
Eq.~(\ref{Eq:Doppler-dispersion}) has the interesting interpretation that the Doppler shift will be large when the interface is approaching (if $v_g>0$) or receding (if $v_g<0$) with a speed $|v| \approx |v_g|$.  This result is exact when the group velocity is constant with frequency.

\subsection{Dispersion and the 1-D Doppler effect on pulse bandwidth}
Intuitively, we expect that a pulse reflecting from an interface whose velocity is close to that of the pulse will be temporally stretched if the interface is receding and compressed if the interface is approaching.  We expect this temporal stretching/compressing should have a corollary decrease or increase of the spectral bandwidth, $\Delta \omega$.  We now show that this is the case.

Adapting expression~(\ref{Eq:gen-bandwidth}), we find
\begin{equation}\label{Eq:Doppler-bandwidth-1}
 \Delta \omega_i-\Delta k_i v=\Delta \omega_r+\Delta k_r v,
\end{equation} 
where $\Delta \omega_i$, $\Delta k_i$, and $\Delta \omega_r$, $\Delta k_r$ are the spectral and spatial bandwidths of the incident and reflected pulses, and $v$ is the interface velocity.  Assuming that the initial and final pulses are sufficiently narrow such that each may be associated with a specific group velocity, we may combine $c v_g=\Delta \omega/\Delta k$ with Eq.~(\ref{Eq:Doppler-bandwidth-1}) to obtain
\begin{equation}\label{Doppler-bandwidth-ratio}
\frac{\Delta \omega_r}{\Delta \omega_i}=
\frac{1-\frac{v}{v_{gi}}}{1+\frac{v}{v_{gr}}},
\end{equation} 
where $v_{gi}$ is the group velocity of the incident pulse and $v_{gr}$ is the group velocity of the receding pulse.
As always, a positive sign for $v$ means a receding interface and a positive sign for $v_g$ means a positive group velocity.  Thus, as expected, when a pulse reflects from an interface that is receding at roughly the pulse group velocity of the approaching pulse, the reflected pulse has a narrowed bandwidth.  When a pulse reflects from a boundary that approaches at roughly the group velocity of the reflected pulse, the reflected pulse is temporally compressed and has a broad bandwidth.

\subsection{How group velocity modulates the frequency response of a pulse to a time-dependent refractive index}
When interface velocity, $v$, is zero, $\phi$ is also zero and $\omega$ is conserved.  When velocity is infinite, $\phi=\pm \pi/2$.  The refractive index is a function of time only and $\vec k$ is conserved.  In this way, the interface model is easily extended to treat the case of a time-dependent (but spatially homogeneous) medium.
Since $k$ is conserved, frequency must compensate for a change in refractive index according to
\begin{equation}\label{Eq:n-omega}
\omega(t)=\frac{n_o}{n(t)}\omega_o,
\end{equation}
where we take the medium to have an initial refractive index $n_o$ and the pulse to have an initial central frequency $\omega_o$.
If $n$ were driven to $0$, $\omega$ would be up-converted by many orders of magnitude.  The changing refractive index associated with plasma generation has been proposed for use in frequency up-conversion in several works \cite{Yablonovitch-1973,Yablonovitch-1974,Wilks-Dawson-Mori-1988,Savage-Joshi-Mori-1992,Berezhiani-Mahajan-Miklaszewski-1999,Geltner-Avitzour-Suckewer-2002,Avitzour-Geltner-Suckewer-2005}.  On a more pedestrian level, frequency response to a time-dependent refractive index is also at the heart of the function of acousto-optic and electro-optic modulators.

Interestingly, the frequency response to a changing refractive index can be modulated by the group velocity of a pulse in a medium.
To see this, we begin with a dispersive medium with a nondispersive time-dependence whose refractive index takes the form $n=n_0(\omega)+t \delta n/\delta t$.  (This might be taken to model, for example, a gas that is a mixture of two atoms, one with resonance that is near $\omega$, the time-dependent frequency of the wave traveling through the medium, and another whose resonances are all spectrally distant but whose concentration changes with time.) Then, from Eq.~(\ref{Eq:n-omega}) we find 
\begin{equation}\label{Eq:ng-modded-response}
 \frac{d\omega}{dt}=\frac{\omega}{n_g}\frac{\delta n}{\delta t},
\end{equation}
where $n_g$ is the group refractive index, defined by $n_g=n+\omega \delta n/\delta\omega$.
One interpretation of Eq. ~(\ref{Eq:ng-modded-response}) is that the frequency response of a pulse to a temporal change in refractive index is proportional to the distance the pulse travels while the change occurs.
\end{section}

\begin{section}{A comparison of two nonstationary homogeneous media: kinetics and dynamics}\label{Sec:comp}

\begin{table*}[htpb]
\centerline{
\begin{tabular}{c l l c c }
Number & Quantity & Symbol & Morgenthaler & Liu\\
\hline
1&  Phase velocity      	&$v_p$ 			&$v_{pf}/v_{pi}$ 	&1 			\\
2&  Group velocity 		&$v_g$ 			&$v_{pf}/v_{pi}$ 	&$v_{gf}/v_{gi}$ 	\\
3&  Permittivity      		&$\epsilon$ 		&$\epsilon_f/\epsilon_i$&1 			\\
4&  Permeability	 	&$\mu$ 			&$\mu_f/\mu_i$ 		&1		 	\\
\hline
5&  Wave number      		&$k$ 			&1 			&1 			\\
6&  Spatial bandwidth    	&$\Delta k$ 		&1		 	&1 			\\
7&  Central frequency 		&$\omega_0$ 		&$v_{pf}/v_{pi}$ 	&1 			\\
8&  Spectral bandwidth 		&$\Delta \omega$ 	&$v_{pf}/v_{pi}$	&$v_{gf}/v_{gi}$	\\
\hline
9&   Total energy density	&$u_t$			&$v_{pf}/v_{pi}$	&1			\\
10&  Electric displacement	&$\vec D$		&1			&$\sqrt{v_{gf}/v_{gi}}$	\\
11&  Magnetic induction		&$\vec B$		&1			&$\sqrt{v_{gf}/v_{gi}}$	\\
12&  Electric field		&$\vec E$		&$\epsilon_i/\epsilon_f$&$\sqrt{v_{gf}/v_{gi}}$	\\
13&  Magnetizing field		&$\vec H$		&$\mu_i/\mu_f$		&$\sqrt{v_{gf}/v_{gi}}$	\\
\hline
14&  Photon energy		&$\hbar \omega$		&$v_{pf}/v_{pi}$	&1			\\
15&  Photon number density	&$N$			&1			&1			\\
16&  Poynting vector		&$S$			&$(v_{pf}/v_{pi})^2$	&$v_{gf}/v_{gi}$	\\
17&  Abraham momentum density	&$\vec E\times \vec H/c^2$&$(v_{pf}/v_{pi})^2$	&$v_{gf}/v_{gi}$	\\
18&  Minkowski momentum density~&$\vec D\times \vec B$	&1			&$v_{gf}/v_{gi}$	\\
19&  Canonical momentum density	&$N(n \hbar \omega/c)$ 	&1			&1			\\
20&  Angular momentum density	&$N(\hbar)$ 		&1			&1			\\
\end{tabular}}
\caption{\label{Tab:comp}A comparison of narrow-pulse transformations by two dynamic media.}
\end{table*}

In this section we seek to augment our understanding of the effects of a time-dependent group velocity in a dispersive medium by comparing them with the effects of a time-dependent \emph{phase} velocity in a \emph{non}dispersive medium.  To do so, we now define two scenarios, which we call Scenarios A and B.  In both scenarios, we imagine a spectrally narrow pulse traveling through a time-dependent medium and ask how that pulse is transformed as the medium changes with time.

In Scenario A, we take the medium to be nondispersive, spatially homogeneous, and isotropic.  We assume that changes in the medium are adiabatic, spatially homogeneous, and isotropic.  We allow $\epsilon$ and $\mu$ to vary gradually and independently over time.  We assume that loss at all times is negligible, meaning that $\epsilon$ and $\mu$ are taken to be real for the narrow spectral range of the pulse.  Because the medium is nondispersive, the group velocity is equal to the phase velocity.

In Scenario B, we take the medium to be dispersive, spatially homogeneous, and isotropic.  We assume that changes in the medium are adiabatic, spatially homogeneous, and isotropic.  We fix $\epsilon$ and $\mu$ for the central frequency, but allow their slopes to vary gradually and independently over time.  We take loss at all times to be negligible, meaning that $\epsilon$ and $\mu$ are taken to be real for the narrow spectral range of the pulse.  

\subsection{Kinetics: basic symmetries}
The spatial homogeneity of Scenarios A and B implies that
\begin{equation}
 \vec k_f=\pm \vec k_i,
\end{equation} and the adiabaticity of the two scenarios eliminates reflection, requiring that we use only the positive sign.  Spatial homogeneity also implies that
\begin{equation}
 \Delta k_f=\Delta k_i
\end{equation}
for the pulses of the two scenarios.
Since $|k|=n\omega/c$ and $\Delta k \approx n_g \Delta \omega/c$, changes in the phase and group velocity imply proportional changes in the central frequency and pulse bandwidth respectively.  Since $n_g=n+\omega\delta n/\delta \omega$, changes in $n$ imply changes in $n_g$ for a nondispersive medium and therefore changes in $n$ lead to changes in bandwidth as well.

To continue our comparison, we now specify particular boundary conditions for the two changing media.  In doing so, we lose the generality of Section~\ref{Sec:symmetry}, but gain the ability to examine the effects of changes to specific, time-dependent media on basic quantities like energy and momentum densities and reflection and refraction coefficients.

We take our two media from previous works.  The non-dispersive medium is taken from Morgenthaler \cite{Morgenthaler-1958}. The dispersive medium is an idealization of the dynamic EIT medium used in the Liu experiment \cite{Liu-Dutton-Behroozi-Hau-2001}.

\subsection{Dynamics: boundary conditions and reflection}
\subsubsection{Morgenthaler's nondispersive medium}
Morgenthaler allowed for the electric permittivity ($\epsilon$) and magnetic permeability ($\mu$) of the medium to be functions of time, but not of space or direction or frequency.  That is,
$\epsilon(\mathbf r, \mathbf k, \mathbf \omega, t)=\epsilon(t)$, and $\mu(\mathbf r, \mathbf k, \mathbf \omega, t)=\mu(t)$.  He assumed that charge and flux, and therefore the electric displacement ($\mathbf D$) and the magnetic induction ($\mathbf B$), would be conserved at an interface boundary.  Under this assumption, the reflection and transmission coefficients for the electric field are given by \cite{Morgenthaler-1958}
\begin{equation}\label{Eq:Er}
 \frac{E_r}{E_i}=\frac{1}{2}\left(\frac{\epsilon_i}{\epsilon_r}-\sqrt\frac{\mu_i\epsilon_i}{\mu_r\epsilon_r}\right),
\end{equation}
and
\begin{equation}\label{Eq:Et}
 \frac{|\mathbf E_t|}{|\mathbf E_i|}=\frac{1}{2}\left(\frac{\epsilon_i}{\epsilon_r}+\sqrt\frac{\mu_i\epsilon_i}{\mu_r\epsilon_r}\right).
\end{equation}
In any Morgenthaler medium, the magnitude of the magnetic field is given by $|\mathbf H|=|\mathbf E|/\eta$, where $\eta=\sqrt{\mu/\epsilon}$ is the impedance.  In general, there is a reflected wave.

We consider the case where changes to the medium are adiabatic and homogeneous.  In this case the reflected power vanishes.  Using the fact that what is preserved across one interface must be conserved across many parallel interfaces (see the beginning of Section~\ref{Sec:symmetry}), we find that we may take the boundary conditions across an adiabatic change to be
\begin{equation}
 \mathbf D_f=\mathbf D_i
\end{equation} 
and 
\begin{equation}
 \mathbf B_f=\mathbf B_i.
\end{equation} 
The values of $\mathbf E$ and $\mathbf H$ at any time can then be found via $\epsilon$ and $\mu$.

\subsubsection{Liu's dispersive medium}
\begin{figure}
	\centerline{\includegraphics[width=8.5cm]{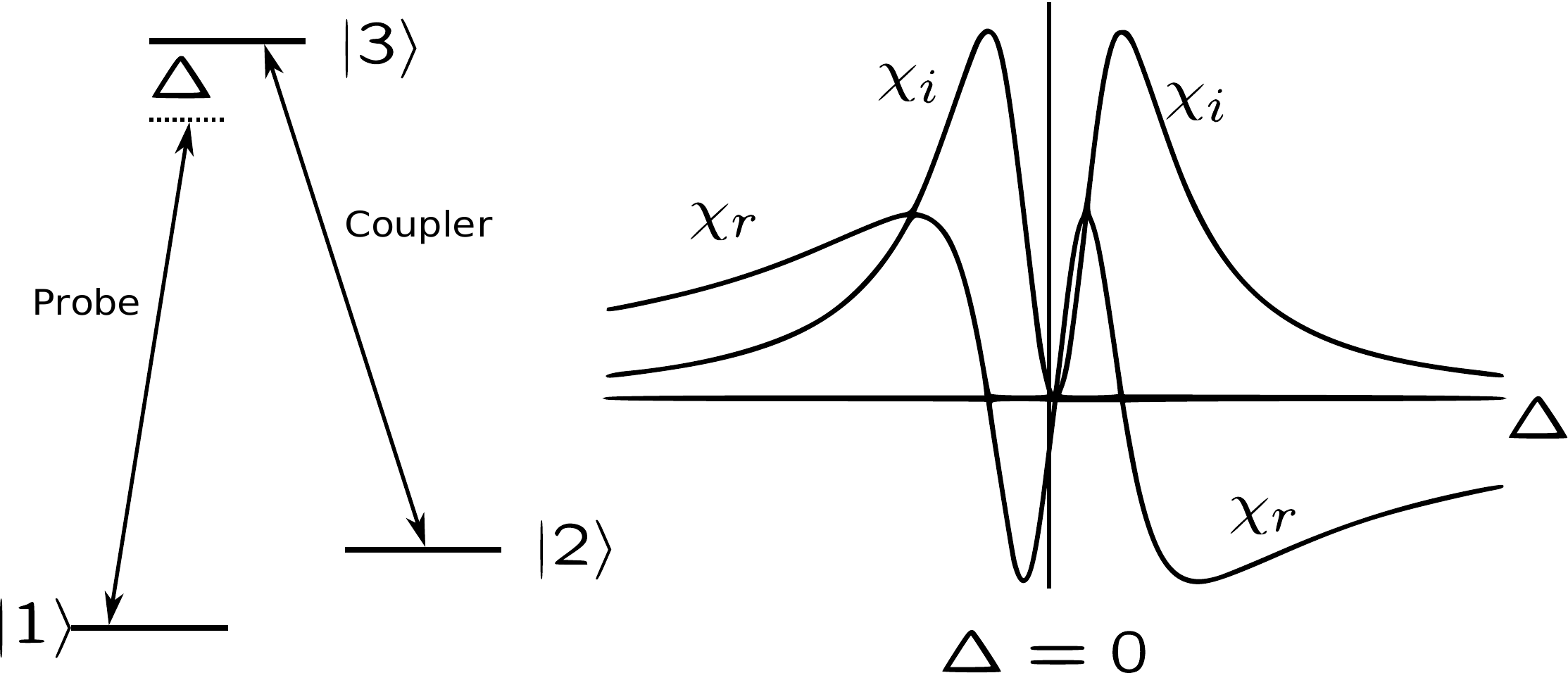}}
	\caption{\label{Fig:EIT2} EIT energy diagram.  The coupling beam and the medium may be taken together to form an effective medium as felt by the probe beam.  The properties of the effective medium depend on the state of the coupling beam.}
\end{figure}
The Liu experiment involves a pulse interacting with an EIT medium across a narrow spectral range around the EIT resonance.  The system as a whole comprises two beams of light interacting with a cloud of cold atoms.  The two beams are tuned to two connected atomic transitions which share an excited state. The stronger beam, referred to as the coupling beam in Figure~\ref{Fig:EIT2}, is considered in combination with the cloud of atoms to form the EIT medium, as seen by the weaker beam, which we call the probe beam. When EIT is established, the atoms are in a coherent dark state, so-called because it does not couple to the probe beam.  The dark state may be represented as a combination of electronic levels $|1\rangle$ and $|2\rangle$ as 
\begin{equation}
 |D\rangle=\frac{\Omega_c |1\rangle-\Omega_p|2\rangle\exp[i (\vec k_p-\vec k_c)\cdot\vec r-i(\omega_p-\omega_c)t]}{\sqrt{\Omega_c^2+\Omega_p^2}}.
\end{equation}
If the probe beam is sufficiently weak such that the atomic number density remains significantly larger than the probe photon number density, then the EIT medium will be approximately linear in probe intensity, so that the only sizable nonlinear effects of the EIT medium are the effects of the coupling beam on the probe beam.  Surprisingly, this linear regime may be maintained even as the coupling beam is completely attenuated \cite{Fleischhauer-Lukin-2000}.

For our nondispersive medium, we idealize the Liu experiment \cite{Liu-Dutton-Behroozi-Hau-2001} by neglecting decoherence and other losses and require that total pulse energy be conserved under changes to the group velocity.  Thus,
\begin{equation}
 u_{tf}=u_{ti},
\end{equation}
where $u_t$ represents the total energy density associated with the wave including not only that portion that is in field form but also that portion that is coherently stored in the medium and is a cycle-averaged quantity.  For a narrow-band pulse propagating through a lossless but dispersive medium, we may define a non-dispersive energy density ($u_n$) which is related to the dispersive energy ($u_t)$ by
\begin{equation}
 u_n=u_t\frac{n}{n_g}.
\end{equation} 
In terms of the macroscopic fields, $u_n$ is given, for the isotropic medium we are considering, by 
\begin{equation}
 u_n=\frac{1}{2}\left(\epsilon E^2+\mu H^2 \right),
\end{equation} 
where $E^2$ and $H^2$ are time averaged quantities.
(Note that $u_n$ matches the standard form for the electromagnetic energy density in a lossless, dispersionless, macroscopic medium \cite{Jackson-1998}.  When dispersion becomes negligible, $n_g\approx n$, and $u_t$ reduces to this form.)

If, as we have assumed for our comparison, changes in the group velocity occur around a constant refractive index, then we may take the quantity $u_n n_g$ to be fixed.  We will also assume that the impedance at line center is unchanged, so that the ratio between $H$ and $E$ remains fixed.  Because the refractive index at line center will also remain fixed, all macroscopic electromagnetic fields scale together.  Therefore, all terms quadratic in any pair of fields scale with the group velocity.  The direction of the fields is unchanged by the homogeneous change in the isotropic medium.  We summarize this as
\begin{equation}
 \mathbf F_f \sqrt{v_{gf}}=\mathbf F_i \sqrt{v_{gi}},
\end{equation} 
where $\mathbf F$ could be any of the four macroscopic electromagnetic fields.
\subsection{Transformations wrought under changing phase and group velocities in the `Morgenthaler' and `Liu' media}

We are now able to perform a more complete comparison between scenarios A and B.  In Table~\ref{Tab:comp}, we list 20 different quantities and show how they are scaled as the media of Scenarios A (Morgenthaler) and B (Liu) change.  Quantities 1-4 summarize the difference between the two scenarios, emphasizing that the phase velocity and group velocity change in Scenario A while only the group velocity changes in Scenario B.  The scaling of quantities 5-8 depends only on symmetry (see Section~\ref{Sec:symmetry}) and therefore are general consequences of the scaling of quantities 1-4.  The scaling of quantities 9-13 may be taken as expressions of the boundary conditions for the two media.  

The remaining quantities (14-20) can be found in simple ways from a knowledge of the scaling of the first 13 quantities.  The per photon energy (14) scales with frequency, which scales as in line 7.  That the photon number density (15) is conserved is related to our assumption of no loss.  The Poynting vector (quantity 16) scales as the square of the phase velocity for the non-dispersive case.  This quadratic dependence combines the linear dependence of the total energy density on the phase velocity with the fact that the group velocity is equal to the phase velocity for a nondispersive medium.  Because the dispersive case assumes that the phase velocity remains unchanged, the Poynting vector of the Liu medium feels only the effect of the change in group velocity.  The Minkowski momentum (18) is conserved in the nondispersive case, but is proportional to the group velocity for the nondispersive case.  This second dependence underscores the fact that the Minkowski momentum, as defined by $\mathbf D\times \mathbf B$, does not give the total momentum when dispersion is taken into account.  The canonical momentum (19), which does give the total momentum, is conserved for both media.  Because the total photon number is conserved for both media, so is the total angular momentum density associated with the pulse for both media.  We note that although the total angular momentum does not change, its distribution between field and medium does.

In concluding this section we emphasize three points.  First, to understand the effects of a particular medium on a pulse, it is necessary to specify appropriate boundary conditions that depend on the details of the way in which the changing medium interacts with the field.  Second, boundary conditions that are reasonable for a non-dispersive medium (Scenario A) may fail for a simple dispersive medium (Scenario B) for the particular reason that changes in dispersion imply a shifting of energy from field to medium.  It then becomes necessary to model different changes on a case-by-case basis.  Finally, a general comparison of the scaling of different quantities for Scenario A and Scenario B shows that the changes wrought in a controllably dispersive nonstationary medium are qualitatively different from those wrought in a nonstationary nondispersive medium.  In this sense, controllably dispersive media open up a new and potentially fruitful niche in nonstationary electromagnetics.
\end{section}
\begin{section}{Conclusion}
Controllably dispersive media are an experimental reality \cite{Wang-Kuzmich-Dogariu-2000,Liu-Dutton-Behroozi-Hau-2001, Ginsberg-Garner-Hau-2007}.  All the requirements necessary to perform interesting nonstationary experiments in controllably dispersive media have been demonstrated for one experimental system \cite{Liu-Dutton-Behroozi-Hau-2001, Ginsberg-Garner-Hau-2007}, are near at hand for another \cite{Wang-Kuzmich-Dogariu-2000}, and appear to be feasible for others \cite{Wicht-Rinkleff-Danzmann-2002,Avitzour-Shvets-2008}.  In this paper we have outlined a basic theory for the treatment of controllably dispersive media from the perspective of nonstationary electromagnetics.  

Although allowing the properties of a medium to depend not only on position but also on both time and frequency opens up a large parameter space, the behavior of waves in this space is subject to basic kinetic constraints imposed by symmetry.  We have explicitly defined the requirements of symmetry for the case of a plane wave interacting with a moving planar interface. We have justified these requirements by rederiving well-established results, such as Snell's Law and the free space Doppler shift, and used them to simply derive analytical descriptions of more esoteric phenomena, such as the effect of dispersion on the reflective Doppler shift.  Using the outlined kinetic theory, we have provided an analysis that corroborates a recent proposal \cite{Avitzour-Shvets-2008} for the compression and decompression of pulse envelopes in magnetized plasmas using strictly nonstationary effects and have shown that the proposed mechanism may be applied in any controllably dispersive medium.  We have introduced a simple expression that shows how a pulse may be significantly compressed or decompressed under Doppler reflection--even if the Doppler shifts themselves are small.  Finally, we have shown that dispersion modulates the frequency response of a wave to a temporal change in the refractive index--the response is damped for slow light media, and amplified for fast light media.

Unlike kinetic effects, which rely only on simple symmetries, dynamic effects are determined by boundary conditions which depend upon particular details of the medium at hand.  We have compared dynamic effects in two simple, homogeneous, adiabatically nonstationary media--one which is nonstationary in phase velocity and a second which is nonstationary in group velocity.  We compared the effective boundary conditions associated with these media and the changes wrought in spectrally narrow pulses propagating through them.  We have found substantial differences in both the boundary conditions and the effect of the nonstationary medium on fundamental quantities such as energy, and electromagnetic momentum.  These differences suggest that it may be useful to consider controllably dispersive media as a new and potentially fruitful category of nonstationary electromagnetic media.

We would like to thank Kevin Mertes, John Mcguire, Karen Tate, Yoav Avitzour, and Peter Milonni for help in the development of this work.  In addition, we express gratitude for a Los Alamos Laboratory Directed Research and Development grant, which funded much of this work.
\end{section}


\end{document}